\documentclass[aps,preprint,superscriptaddress,floatfix]{revtex4-1}
\usepackage{graphicx}
\usepackage{amsmath}
\usepackage{amsfonts}
\usepackage{amssymb}
\usepackage{epsfig}

\begin{document}

\title{Decoherence and relaxation of a single electron in a one dimensional conductor}

\author
{A. Marguerite$^{1}$, C. Cabart$^{1,2}$, C. Wahl$^{3 }$, B. Roussel$^{2}$, V. Freulon$^{1 }$, D. Ferraro$^{2,3}$, Ch. Grenier $^{2}$, J.-M. Berroir$^{1 }$, B. Pla\c{c}ais$^{1 }$, T. Jonckheere$^{3 }$, J. Rech$^{3 }$, T. Martin$^{3 }$, P. Degiovanni$^{2}$, A. Cavanna$^{4 }$, Y. Jin$^{4 }$, and G. F\`{e}ve$^{1 \ast}$ \\
\normalsize{$^{1}$Laboratoire Pierre Aigrain, Ecole Normale Sup\'erieure-PSL Research University, CNRS, Universit\'e Pierre et Marie Curie-Sorbonne Universit\'es, Universit\'e Paris Diderot-Sorbonne Paris Cit\'e, 24 rue Lhomond, 75231 Paris Cedex 05, France}\\
\normalsize{$^{2}$ Univ Lyon, ENS de Lyon, Univ Claude Bernard, CNRS,
Laboratoire de Physique, F-69342 Lyon, France} \\
\normalsize{46 All\'ee d'Italie, 69364 Lyon Cedex 07,France.}\\
\normalsize{$^{3}$  Aix Marseille Universit\'e, Universit\'e de Toulon, CNRS, CPT, UMR 7332, 13288 Marseille, France}\\
\normalsize{$^{4}$Laboratoire de Photonique et Nanostructures (LPN)}\\
\normalsize{CNRS, Universit\'{e} Paris-Saclay, Route de Nozay, 91460 Marcoussis, France}\\
\normalsize{$^\ast$ To whom correspondence should be addressed;
E-mail:  feve@lpa.ens.fr.}
}

\begin{abstract}
We study the decoherence and relaxation of a single elementary electronic excitation propagating in a one-dimensional chiral conductor. Using two-particle interferences in the electronic analog of the Hong-Ou-Mandel experiment, we analyze quantitatively the decoherence scenario of a single electron propagating along a quantum Hall edge channel at filling factor 2. The decoherence results from the emergence of collective neutral excitations induced by Coulomb interaction and leading, in one dimension, to the destruction of the elementary quasiparticle. This study establishes the relevance of electron quantum optics setups to provide stringent tests of strong interaction effects in one-dimensional conductors described by Luttinger liquids paradigm.
\end{abstract}

\maketitle

\section{Introduction}

What is the fate of a single electron propagating in a conductor? This basic problem of condensed matter physics has given birth to the Landau-Fermi liquid paradigm in 3D solids: Coulomb interactions limit the phase coherence at low temperatures \cite{Pierre2003} but nevertheless do not invalidate the single electron as a quasiparticle \cite{Pines1966}. At the opposite in 1D systems, Coulomb interactions favor the emergence of collective modes \cite{Luttinger, Auslaender} leading to the disappearance of the single electron as a good quasiparticle \cite{LeHur:2002-1, Ferraro2014}, thus giving rise to the Luttinger liquid paradigm for 1D quantum wires \cite{Haldane1981}. However, observing this transition from single to many body physics requires a more elaborate experimental scheme than a direct measurement of the current flowing in the conductor as charge propagation remains unaffected by interactions.
Coulomb interactions have already been shown to be responsible for electronic decoherence \cite{Neder2006, Roulleau2008, Tewari2015} in single particle interferometers \cite{Ji2003, McClure2009} and for the relaxation of non-equilibrium energy distribution \cite{Altimiras2010, LeSueur:2010-1}. However the fate of the single elementary quasiparticle could not be accessed as stationary sources were used, lacking both temporal dependence and single electron resolution. In this paper, we use two-particle interferences in the electronic analog \cite{Ol'khovskaya2008, Bocquillon2013} of the Hong-Ou-Mandel (HOM)\cite{HOM} experiment to analyze quantitatively the decoherence \cite{Wahl2014} of a single electron along its propagation within the outer edge channel of the integer quantum Hall effect at filling factor 2. The HOM experiment thus probes for the first time the decay of a single Landau quasiparticle in a ballistic conductor. By revealing the fate of a single electronic excitation and confirming our predictions for its decoherence scenario\cite{Wahl2014, Ferraro2014}, this study demonstrates how electron quantum optics techniques provide a powerful probe of strong interaction effects in ballistic conductors.

The characterization of a single electron state stems from the study of its coherence properties. The single electron coherence of an electron source at position $x$ of an edge channel \cite{Grenier2011a, Haack2011} can be defined using a quantum optics formalism \cite{Bocquillon2014}: $\mathcal{G}_x^{(e)}(t,t') = \langle \Psi^{\dag}(x, t')
\Psi(x, t) \rangle $, where the electric field of quantum optics has been
replaced by the fermion field operator $ \Psi(x, t)$ which annihilates an
electron at time $t$ and position $x$ of the edge channel (as the position $x$
will be fixed, it will be dropped in the following). Most experiments investigating electron coherence have been performed using stationary sources (using a dc voltage bias) which
continuously emit electrons in the conductor. In the stationary case, $\mathcal{G}^{(e)}$ only depends on the time difference $\tilde{\tau}=t-t'$ and provides information on the coherence time of the source, but does not depend on $\bar{t}= (t+t')/2$ . Single electron coherence is thus fully determined by the mere knowledge of the energy distribution of the emitted electrons (through Fourier transformation).  The situation is very different when one
uses the recently developed single electron emitters \cite{Feve2007, Fletcher2013, Dubois2013, Ubbelohde2015, Kataoka2016}, which trigger the
emission of a single particle at a well defined time, such that the full $t,
t'$ dependence needs to be retained. For an ideal single electron source, $\mathcal{G}^{(e)}(t,t')$
fully encodes  the emitted single electron wavepacket  $\varphi_e(t)$:
$\Delta \mathcal{G}^{(e)}(t,t') = \mathcal{G}^{(e)}(t,t') - \mathcal{G}_F^{(e)}(t-t') = \varphi_e^{*}(t')
\varphi_e(t)$, where $\mathcal{G}_F^{(e)}$ is the equilibrium contribution of the Fermi
sea.  Probing single electron coherence thus brings the possibility to picture single particle states
propagating in solid state \cite{Grenier2011a, Jullien2014} and characterize how single electron wavepackets evolve under the influence of Coulomb interaction. In this non-stationary situation, in the frequency domain, $\widetilde{\mathcal{G}}^{(e)}(\omega,\omega')$ has two Fourier components $\omega$ and $\omega'$ \cite{Grenier2011a}. The energy distribution, which has been recently measured \cite{Altimiras2010, LeSueur:2010-1}, only contains a partial information on single electron coherence (diagonal part $\omega=\omega'$). The non-diagonal elements ($\omega \neq \omega'$) contain all the information on the non-stationary aspects of electron coherence ($\bar{t}$ dependence). As we shall see, their knowledge is thus essential for predicting HOM traces but also to predict the evolution of the non-stationary state under the influence of Coulomb interaction. In the presence of strong Coulomb interactions, this problem cannot be reduced to the redistribution of the electron energy distribution caused by Coulomb induced electron-electron scattering.

This paper is organized as follow. In Section II we discuss the properties of single electron coherence and how it can be efficiently probed in the Hong-Ou-Mandel experiment. Experimental results are presented in Section III. Single electron decoherence shows up in the electronic HOM experiment as a reduction of the two-particle interference contrast. These results are discussed using a phenomenological approach of electronic decoherence. In Section IV we present various possible mechanisms which could lead to the observed contrast reduction. By ruling out progressively the majority of them by a direct comparison with the experimental results, we can identify the inter-channel Coulomb interaction as the dominant source of contrast reduction. In Section V, we quantitatively compare our experimental observations with non-perturbative bosonization-based models for
Coulomb interaction induced single electron decoherence at filling factor $\nu=2$. This allows us to properly fit the experimental HOM traces without the need to introduce any unnecessary phenomenological parameters. This agreement establishes the complete picture of the destruction scenario of a single electron in a chiral one-dimensional conductor.

\section{Electron coherence in the electronic Hong-Ou-Mandel experiment}

\begin{figure}[hhhhhh]
\centerline{\includegraphics[width=\columnwidth]{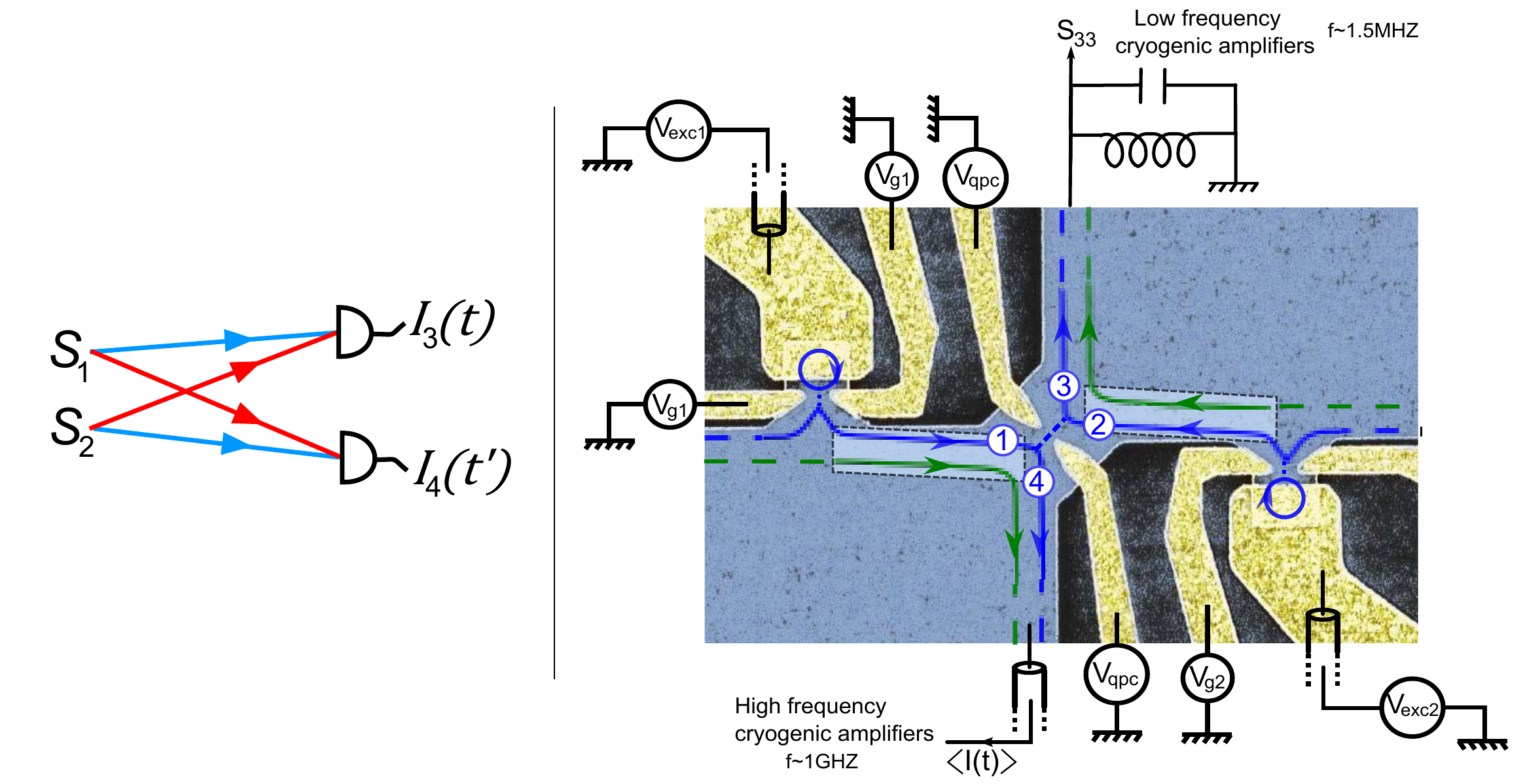}}
 \caption{ \textbf{Hong-Ou-Mandel interferometry} Left panel, sketch of two-particle interferences. Right panel, modified scanning electronic microscope picture of the sample. The electron gas is represented in blue, the edge channels by blue (outer channel) and green (inner channel) lines, metallic gates are in gold. The emitters are placed at inputs 1 and 2 of the QPC used as an electronic beam-splitter. Single electron emission by source $i$ on the outer channel is triggered by the square voltage $V_{\text{exc}, i}$ of amplitude $0.7$ K. The dot to edge transmission of source $i$ is tuned by the gate voltage $V_{\text{g}, i}$. The central QPC is set to partition ($R=0.5$) the outer channel using the gate voltage $V_{\text{qpc}}$. Interaction regions of length $l\approx 3 \mu m$ are represented by light blue boxes. Average ac current measurements are performed on the splitter output $4$ in order to characterize the source parameters (in particular $\tau_e$). Low frequency noise measurements $\Delta S_{33}$ are performed on output $3$. } \label{Fig1}
\end{figure}

In the electronic HOM experiment (see Fig.\ref{Fig1}, right panel), two quantum dots driven by step voltage pulses (with a repetition frequency $f= 0.9$ GHz and a typical risetime of
30 picoseconds) are used as single electron sources \cite{Feve2007}. The peak to peak voltage pulse  amplitude matches the dot addition energy $\Delta =1.4$ K such that a single electron followed by a single hole are emitted in the outer edge channel at energy $\omega_e=\omega_h=0.7$ K above (electron) or below (hole) the Fermi level. The dot to edge channel transmission can be tuned to vary the escape time
$\tau_e$ (and hence the wavepacket length) of the emitted electrons and holes. The electron sources are placed at $l=3$ $\mu m$ upstream of the inputs 1 and 2 of a quantum point contact (QPC) used as an electronic beam-splitter.

When two electrons collide synchronously on the splitter, two paths (see Fig. \ref{Fig1}, left panel) contribute to the coincidence detection of the detectors placed at outputs 3 and 4 : either a particle in input 1 is detected in output 3 and 2 goes to 4 (blue path) or 1
goes to 4 and 2 goes to 3 (red path). The sum of the two probability
amplitudes for the corresponding exchanged paths leads to interferences
involving two particles at the input \cite{Liu1997, Samuelsson2004, Neder2007}. These two-particle interferences can be revealed by the measurement of the low-frequency current correlations (or noise) at the output of the splitter $\Delta S_{33}$ (where $\Delta$ refers to the excess noise after subtraction of the equilibrium noise contribution), and used to provide information on the excess single electron coherence $\Delta \mathcal{G}^{(e)}_i$ of the sources placed at inputs $i=1,2$ \cite{Grenier2011a}:
\begin{widetext}
\begin{eqnarray}
\Delta S_{33}&=&\Delta S_{\text{HBT}} -\Delta S_{\text{HOM}}  \label{DeltaQ}  \\
\Delta S_{\text{HBT}} & = & 2 R(1-R) \; e^2 f
\int_{0}^{\infty} d\epsilon \big( \delta n^{e/h}_{1}(\epsilon)
+ \delta n^{e/h}_{2}(\epsilon) \big) (1-2f_{\mu,T}(\epsilon)) \label{DeltaSHBT} \\
\Delta S_{\text{HOM}} & =& 4  R(1-R) \; e^2
\int_{-\infty}^{+\infty}d\tilde{\tau} \; \overline{\Delta \mathcal{G}_1^{(e)}(\bar{t}+\frac{\tilde{\tau}}{2} ,\bar{t}-\frac{\tilde{\tau}}{2} ) \;
\Delta \mathcal{G}_2^{(e)}(\bar{t}-\frac{\tilde{\tau}}{2} , \bar{t}+\frac{\tilde{\tau}}{2}  ) }^{\bar{t}} \label{DeltaSHOM}
\end{eqnarray}
\end{widetext}

where $R$ is the reflection probability of the beam-splitter, $f_{\mu,T}(\epsilon)$ the Fermi distribution and $\delta n^{e/h}_{i}(\epsilon)$ the energy density of electron and hole
excitations emitted by source $i$ in one period $1/f$. The first term labeled $\Delta S_{\text{HBT}}$ in (\ref{DeltaQ}) represents the random partitioning of quasiparticles on the beam splitter. As can be seen from Eq.(\ref{DeltaSHBT}), at zero temperature it is directly proportional to the total number of excitations (electrons and holes) emitted by sources 1 and 2. At finite temperature, the partitioning of low energy electron/hole pairs is reduced by two-particle interferences with thermal Fermi sea excitations \cite{Bocquillon2012}. The second term labeled $\Delta S_{\text{HOM}}$ in (\ref{DeltaQ}) is the
two-particle interference term. It comes with a minus sign, as a consequence of the fermionic statistics, and thus reduces the random partitioning.  As can be seen from Eq.(\ref{DeltaSHOM}), it is given by the overlap between the single electron coherence of the two sources and thus probes both their diagonal ($t=t'$ or $\omega= \omega'$ in Fourier space) as well as off-diagonal ($t\neq t'$ or $\omega \neq \omega'$) elements. In the case of pure single electron states
$\varphi_{1,2}(t)$ emitted by each source above the thermal excitations
of the Fermi sea, the general expressions (\ref{DeltaQ}) and (\ref{DeltaSHBT})
have a simplified form and the normalized HOM noise $\Delta q=\Delta S_{33}/ \Delta
S_{\text{HBT}}$ reads:
\begin{eqnarray}
\Delta q & = & 1-\left|\int
\varphi_{1}(t)\varphi^*_{2}(t)\,\mathrm{d}t\right|^2\,.
\end{eqnarray}
The overlap between the states can be experimentally
varied by tuning the emission delay $\tau$ between the two electron sources. For long time delays, classical random partitioning is observed: $\Delta
q = 1$. For short time delay, two-particle interferences occur leading to a dip
in the output noise, the width of which corresponds to the duration of the emitted
wavepackets. Measuring the noise suppression thus provides a quantitative
measurement of the coherence of a single elementary excitation.

HOM interferometry offers several advantages compared to single particle interferometry such as Mach-Zehnder interferometry (MZI), even though it involves the measurement of current correlations instead of current.
First, as noticed by Hanbury Brown and Twiss in their seminal experiment \cite{HBT}, intensity interferometry (such as HOM) is not sensitive to phase fluctuations. Consequently, path lengths only need to be controlled at the
wavepacket scale compared to the wavelength in MZI. This has led to the success of intensity interferometry
in astronomy \cite{HBTb} and, in the present context, enables us to escape the need for controlling the design of the interferometer down to the Fermi wavelength size (few nanometers) as required by amplitude interferometry tomography
protocols \cite{Haack2011}. More importantly, due to its extended nature (few microns), Coulomb interaction is known to occur within the MZI itself \cite{Roulleau2008}. Because the experimental data cannot be corrected for the associated decoherence effects within the MZI, one cannot reconstruct the single electron coherence at the input of the interferometer from the outcoming current measurements. Moreover, predictions for the output coherence can even be hard to obtain. The recent experiment \cite{Tewari2015} in which energy resolved electronic excitations are continuously injected shortly before a MZI illustrates this point. When such an excitation has only been weakly affected along its propagation between its injection point and the input of the MZI, predictions for the output signals can be made \cite{Tewari2015, Slobodeniuk2016, Levkivskyi2008} that compare favorably to the experiment. But the general case where the input state has already been strongly affected by the Coulomb interaction is much more complex and no general prediction has been obtained yet. Such problems do not occur when using the HOM interferometer as it consists of a point-like beam-splitter within which interaction effects can be neglected.

In the following, HOM interferometry is used to quantitatively analyze the decoherence of a single electronic excitation propagating along the outer
edge channel of the integer quantum Hall regime at filling factor $\nu=3$ and $\nu=2$.

\section{Single electron decoherence} \label{sec3}

Figure \ref{Fig2} presents $\Delta q(\tau)$ at filling factor $\nu=2$  for three values of the source emission
time $\tau_e $ which sets the temporal size of the emitted wavepackets.  A dip is
observed for the three curves around $\tau=0$, revealing the reduction of
random partition noise by two-electron interferences. As expected, the width of
the dip increases for increasing escape time $\tau_e$ corresponding to
increasing wavepacket width. However, none of the dips reaches the full
suppression $\Delta q(0) = 0$ showing that we fail to reach the collision of
perfectly indistinguishable wavepackets. Defining the contrast $\gamma$ of two-particle interferences in relation with the depth of the HOM dip: $\gamma= 1- \Delta q(0)$, $\gamma$ is reduced when $\tau_e$ increases.
Electron emission can then be  modeled as a Breit-Wigner resonance between the dot and the edge channel in energy space \cite{Grenier2011a}:
\begin{equation}
\widetilde{\varphi}_{e} (\omega) =\frac{\mathcal{N}_e\Theta(\omega)}{\omega-\omega_e+\frac{i}{2\tau_e}}, \label{eqphifour}
\end{equation}
where $\Theta(\omega)$ is the Heaviside step function
confining the wavepacket to the accessible electronic energy levels
and $\mathcal{N}_e$ a normalization constant. The wavepacket is parametrized by  the
emission energy of the electron $\omega_e=0.7$~K and the resonance width $1/\tau_e$.
The time $\tau_e$ defined in this way indeed corresponds to the wavepacket duration expressed in the time
domain~\cite{Jonckheere2012b}:
\begin{equation}
\varphi_{e} (t) = \frac{\Theta(t)}{\sqrt{\tau_e}}\,e^{i
\omega_{e} t} e^{-t/2\tau_e}. \label{eqphi}
\end{equation}
Without decoherence, such single electron wavepackets would lead
to the following normalized HOM noise:
$\Delta q (\tau)= 1 - e^{-|\tau|/\tau_e}$. To account for the observed non unit contrast $\gamma$, data are fitted (solid lines) by $\Delta q(\tau) = 1 - \gamma e^{-|\tau|/\tau_e}$. The escape times extracted  from the fits
correspond to $75 \pm 6$ (red trace), $110 \pm 13$ (blue trace) and $140 \pm 13$
ps (black trace). They differ, in particular for the shortest time, from the measurements of $\tau_e$ extracted from the phase of the a.c. current first harmonic generated by the sources\cite{Feve2007}: $\tau_e=30 \pm 5$, $100 \pm 18$ and $180 \pm 50$ ps. The difference can be understood firstly by the non-zero rise time of the excitation pulse (typically $30$ ps) and secondly by the inter-channel Coulomb interaction which leads to a widening of the current pulse \cite{Freulon2015}. Both effects are not accounted for by our independent measurement of $\tau_e$ from the a.c. current phase which probes the time delay between the excitation pulse and the emitted current. It is thus neither sensitive to the excitation pulse rise time nor to the charge fractionalization occurring on the outer channel \cite{Freulon2015} as the current measurement only probes the charge mode. This explains the data/fit discrepancy for short emission times. For long emission times, the relative importance of these effects decreases and the agreement with the exponential description is better. The contrast $\gamma$ extracted from the fits decreases from $0.65 \pm 0.04$ (red line) to $0.43 \pm 0.035$ (blue line) and $0.3 \pm 0.015$ (black line) showing that the indistinguishability of incoming electrons decreases with
increasing escape time. On long escape times, two-electron interferences are almost fully suppressed.

\section{Decoherence scenarii}
Several possibilities could be envisioned. A first hypothesis is that the emission of undistinguishable electrons is prevented by
differences between the two sources, either from sample construction or related to environmental noise leading to random
fluctuations  in the energy at which electrons are emitted~\cite{Kato2014}. In a second hypothesis, the contrast reduction could be related to an energy dependence of the  beam-splitter reflection $R(\epsilon)$ \cite{Marian2015}. However, according to \cite{Marian2015}, the contrast should increase with increasing wavepacket length which is opposite to what we observe.  Finally, the third hypothesis relies on Coulomb induced decoherence~\cite{Wahl2014,Ferraro2014}
along propagation as the quasiparticle gets entangled with the neighboring edge channel acting as an environment~\cite{Freulon2015}. As discussed in the introduction, we shall see now that a careful analysis of the experimental data supports the Coulomb induced decoherence scenario.

\begin{figure}[hhhhhh]
\centerline{\includegraphics[width=12cm]{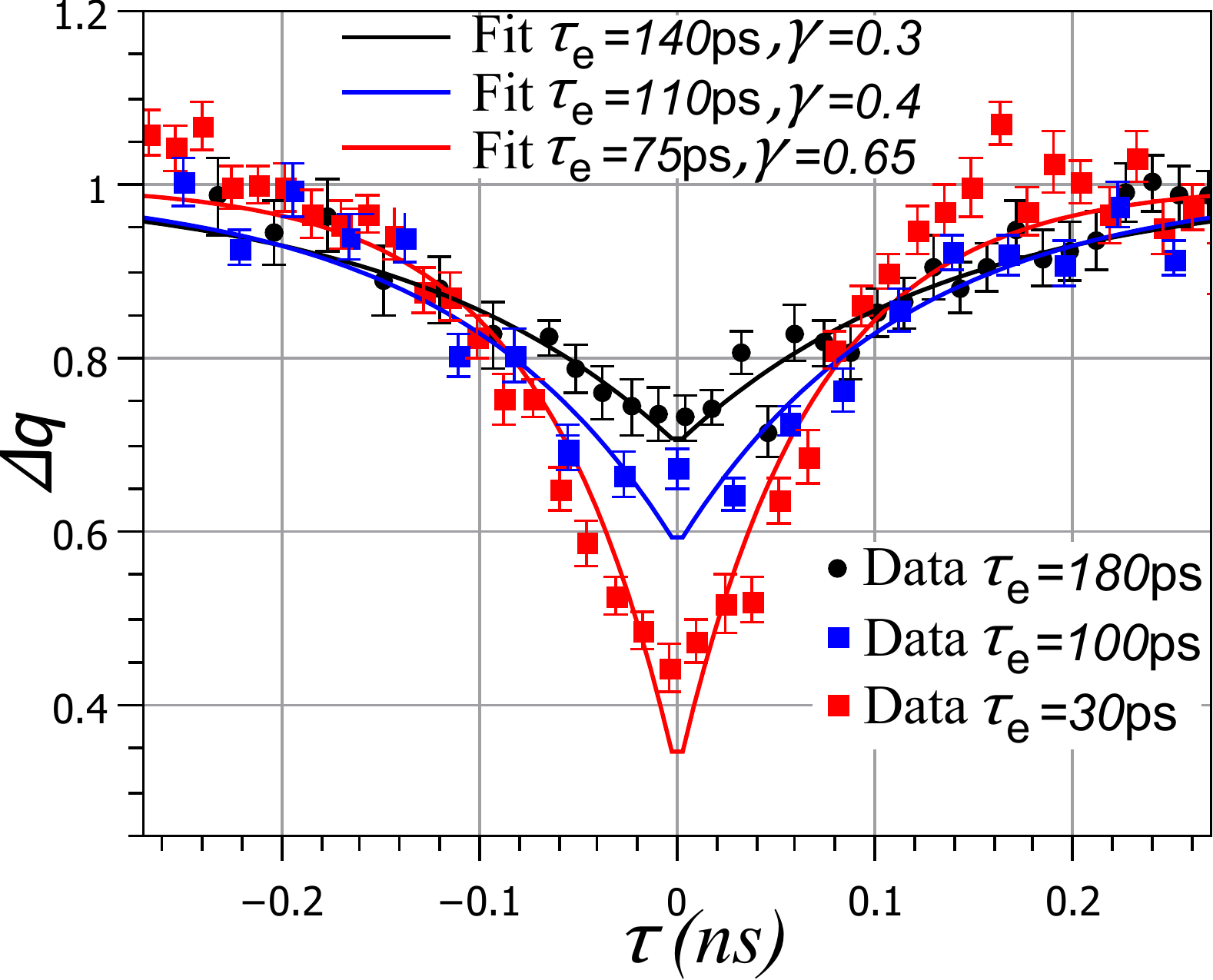}}
 \caption{ \textbf{Electronic Hong-Ou-Mandel experiment.} HOM trace $\Delta q$ as a function of the time delay between the sources $\tau$ for three values of the emission time $\tau_e = 30$ ps (red squares), $\tau_e = 100$ ps (blue squares) and $\tau_e = 180$ ps (black dots). The plain lines represent exponential fits, $\Delta q(\tau) =1-\gamma e^{-|\tau|/\tau_e}$.}
 \label{Fig2}
\end{figure}

\begin{figure}[hhhhhh]
\centerline{\includegraphics[width=12cm]{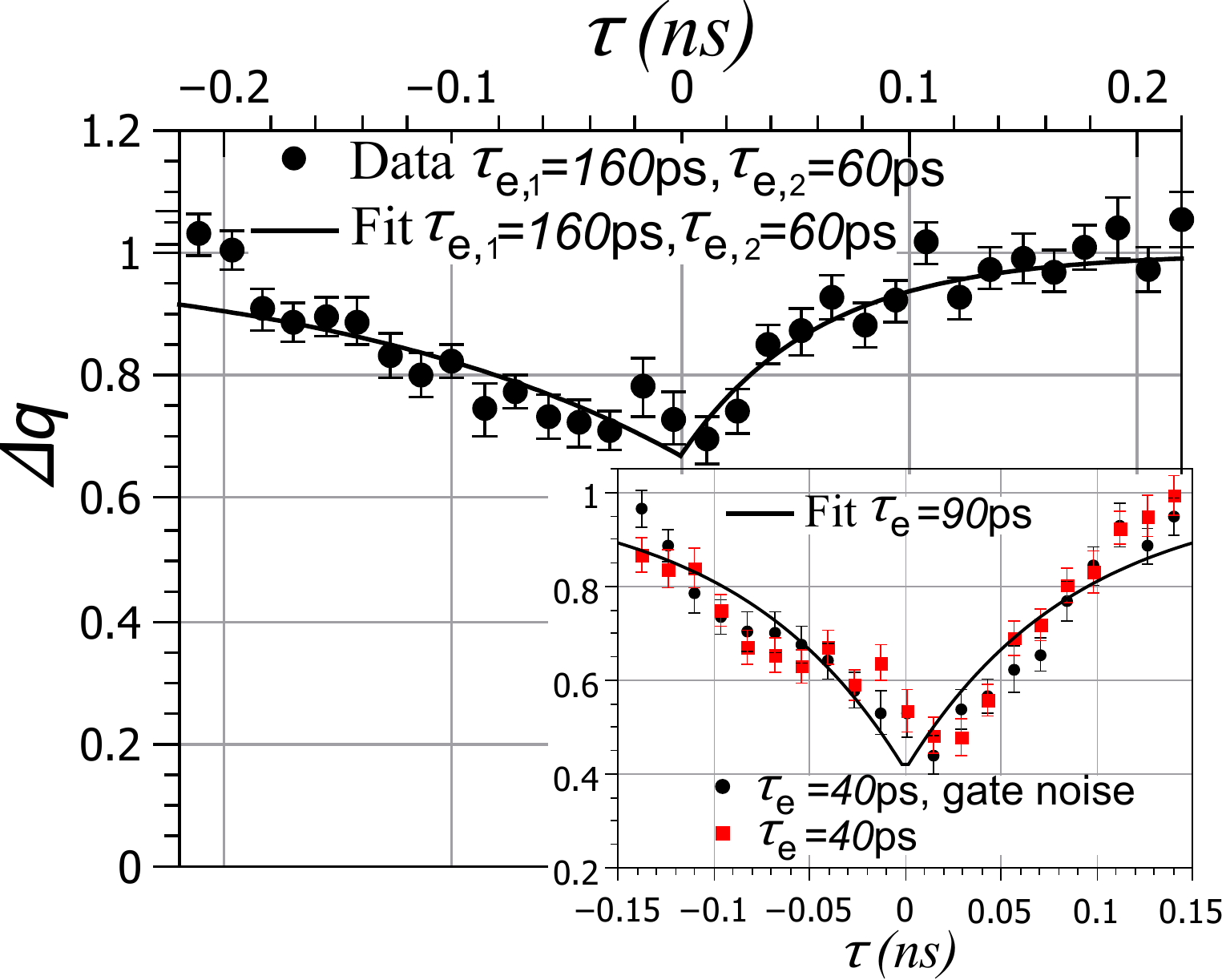}}
 \caption{ \textbf{Escape time asymmetry and energy emission fluctuations.} $\Delta q(\tau)$ for asymmetric ($\tau_1=160$, $\tau_2=60$ ps, black dots) escape times. The black line is an exponential fit with $\tau_e= 160$ ps for $\tau \leq 0$, $\tau_e= 60$ ps for $\tau \geq 0$. Inset, $\Delta q(\tau)$ with (black dots) and without (red squares) external noise applied on the static potential of dot 1. The noise amplitude corresponds to a blurring of a 400 mK of the dot emission energy. The black line is a fit with an exponential decay. } \label{Fig3}
\end{figure}

Considering the first
hypothesis, there are two parameters on which we can act to tune the source:
the escape time $\tau_e$ by varying the potential barrier coupling the dot to
the edge channel, and the energy $\omega_e$ at which electrons
are emitted by changing the dot static potential. As can be seen on Fig.~\ref{Fig3} the effect of escape time
detuning ($\tau_{e,1}=60$ ps, $\tau_{e,2}=160$ ps) can easily be seen on the HOM trace. The curve $\Delta q(\tau)$
becomes asymmetric (see black trace), as
predicted theoretically~\cite{Jonckheere2012b}:
one side falls very quickly (source with short escape time)
while the other falls much more slowly (source with long escape time). The observed contrast is very close to the one measured for $\tau_{e,1}=\tau_{e,2}=100$ ps (see Fig.\ref{Fig2}) corresponding roughly to the average emission time in the detuned case. Escape time asymmetries thus cannot explain the contrast reduction observed on Fig.~\ref{Fig2}.
Surprisingly, no variation of the contrast is observed when the emission energies are detuned by varying the dot
potential or by applying on purpose external noise to the static dot
potential (see inset of Fig.~\ref{Fig3} where a blurring of approximately $400$~mK of the emission energy of one dot is applied).
The effects of energy fluctuations can be estimated theoretically by averaging HOM traces calculated using the Floquet scattering formalism \cite{Moskalets2002, Parmentier2012, Moskalets2013} using gaussian fluctuations of amplitude $k_b T_n$ of the energy of one dot. It predicts that for $\tau_e=40$ ps, the contrast should vary from $0.8$ to $0.5$ when the noise amplitude varies from $T_n=140$ mK to $T_n=400$ mK. Here the value $T_n=400$ mK corresponds to the amplitude of the applied gate noise on Fig.\ref{Fig3} (insert) while $T_n=140$ mK is the maximum spurious gate noise compatible with the width of the conductance peaks deduced from current measurements as a function of the dot potential. On the contrary, we observe no substantial modification of the contrast when the additional noise is applied (whereas an almost complete blurring of the conductance peaks is observed). This means that our experimental data do not support the inhomogeneous broadening hypothesis.

\begin{figure}[hhhhhh]
\centerline{\includegraphics[width=12cm]{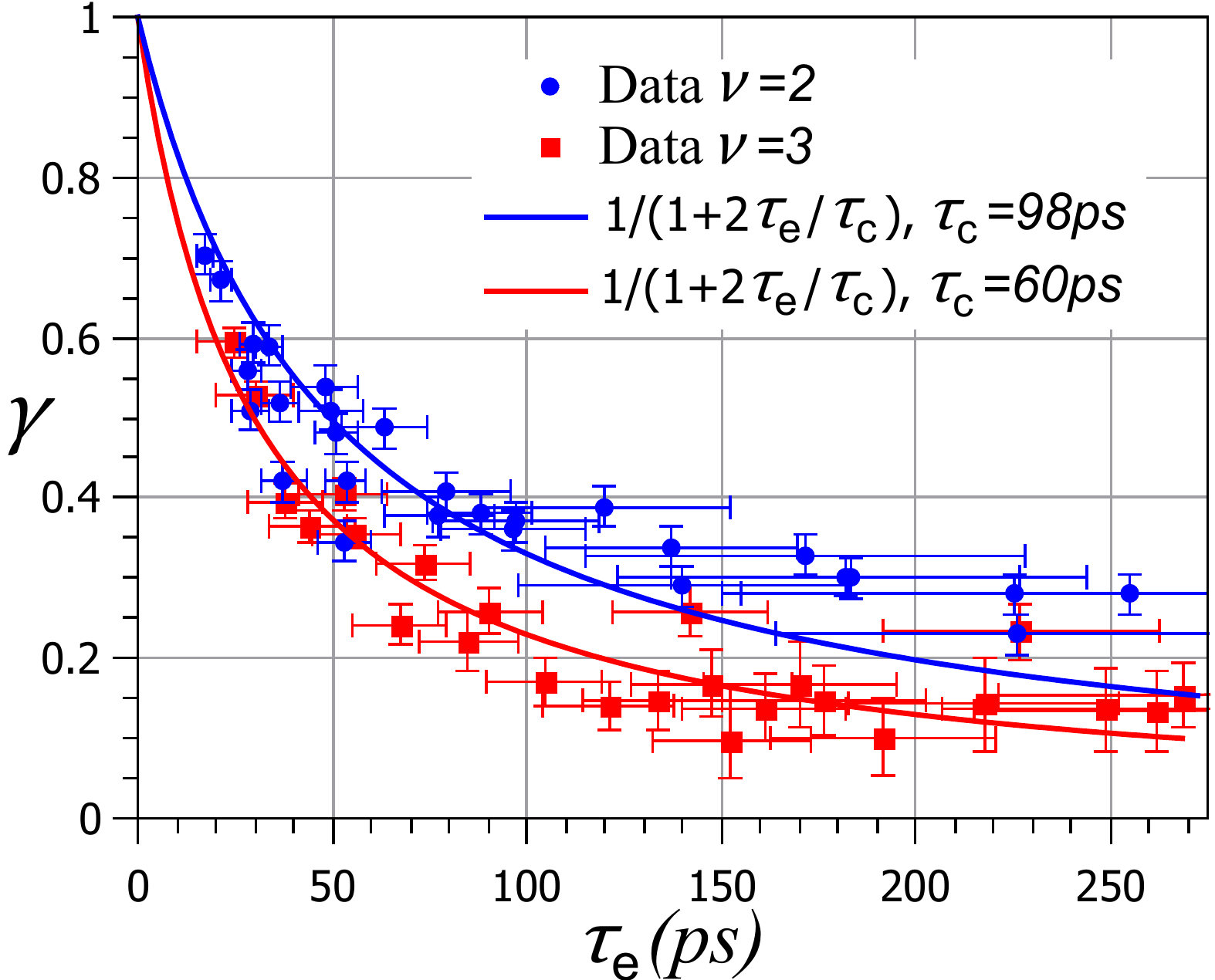}}
 \caption{ \textbf{Contrast versus emission time.} Evolution of contrast $\gamma$ as a function of emission time $\tau_e$ for $\nu=2$ (blue dots) and $\nu=3$ (red squares). The plain lines correspond to the fits by the phenomenogical model $\gamma (\tau_e) =1/(1+ 2 \tau_e/\tau_c)$. } \label{Fig4}
\end{figure}

To understand the
mechanism responsible for the contrast reduction, we have plotted the variation
of contrast $\gamma$ with emission time $\tau_e$ for filling factors 2 and 3 on
Fig.~\ref{Fig4}: it falls quickly on short
time both for $\nu=2$ and $\nu=3$ with  faster contrast
reduction for $\nu=3$ compared to $\nu=2$. To account for the contrast
dependence in the escape time, let us introduce a phenomenogical coherence
time $\tau_c$ on which the off diagonal terms of the coherence
decay to zero : $\Delta \mathcal{G}^{(e)}(t,t') \rightarrow
e^{-|t-t'|/\tau_c}
\Delta\mathcal{G}^{(e)}(t,t')$.  Then, only time components $(t,t')$ with
$|t-t'|\leq \tau_c$ of the wavepacket can interfere on the splitter
whereas
$|t-t'|\geq \tau_c$ components are subject to random partitioning.
 Using \eqref{DeltaQ}, this phenomenological decoherence factor predicts a reduction of the contrast related to the ratio $\tau_c/\tau_e$: $\gamma= (1+2 \tau_e/\tau_c)^{-1}$ where the factor
2 reflects the presence of two sources on which decoherence acts.
The plain lines on Fig.~\ref{Fig4} represent fits with the above mentioned expression
$\gamma(\tau_e)$, providing $\tau_c=60$ ps at $\nu=3$ (red line) and $\tau_c = 98$ ps at $\nu=2$ (blue line).
The difference in $\tau_c$ between $\nu=3$ and
$\nu=2$ suggests that decoherence occurs during propagation and is related to
interchannel Coulomb interaction which depends on the number of
co-propagating edge channels.

\section{Single electron fractionalization: the death of the elementary quasiparticle}

Let us now review the theoretical models for Coulomb interaction effects on single
electron coherence along propagation within two copropagating
edge channels. Intra and inter-channel effective screened Coulomb
interaction can be discussed efficiently within the bosonization \cite{vonDelft1998}
framework. It essentially states that all
excitations of a one-dimensional chiral edge channel can be
described in terms of collective bosonic modes, called edge magnetoplasmons (EMP). At $\nu=2$, strong interactions between the two channels lead to the emergence of the bosonic symmetric charge eigenmode with velocity $v_{\rho}$ and the anti-symmetric neutral or dipolar eigenmode with velocity $v_n$ \cite{Levkivskyi2008}. Since its introduction, this
physical picture has been directly confirmed, at least at low enough frequencies (below 10
GHz) through finite frequency admittance measurements \cite{Bocquillon2013b}. As the two modes propagate at different
velocities, a single electronic excitation propagating on length $l$ in the outer
channel splits into two pulses carrying a fractional charge~\cite{Berg2009, Freulon2015}
$e/2$ separated by the fractionalization time $\tau_s=l/v_n - l/v_{\rho}$.
This phenomenon already discussed in
the context of quantum wires (1D non-chiral Luttinger
liquids)~\cite{Safi1995, Steinberg2007, Kamata2014} is expected to induce
the disappearance of the electron as a genuine
quasiparticle~\cite{LeHur:2002-1}.

We now turn to the modelization of Coulomb interaction in our experiment. Following Ref.\cite{Ferraro2014}, we decompose the edge channel into three parts. The first is a non-interacting region located at $x=0$ in which the electron injection occurs. We assume here that, as experimentally observed \cite{Feve2007, Parmentier2012}, the tunneling process from the dot is not affected by interactions apart from the renormalization of the dot parameters (electron emission in a strongly interacting system like the fractional quantum Hall effect deserves a different discussion \cite{Ferraro2015}). The second is the interaction region ($0<x<l$) of length $l\approx 3 \mu m$ where we describe the Coulomb interaction as a local density-density interaction term acting within each edge channel (intrachannel interaction) and between edge channels (interchannel interaction) and fully parameterized by the fractionalization time $\tau_s$. The third part is the beam-splitter where we assume that electrons are locally non-interacting and that Coulomb interaction does not couple together edge channels located on opposite sides of the splitter. Importantly, the chirality of the edge channels as well as the short-range nature of the interaction (which constitutes a good approximation at low enough energy \cite{Bocquillon2013b}) allow us to extend the interaction region from infinitesimally after the point of injection to infinitesimally before the beam splitter and to neglect any back-action of the interaction region on the emission process from the quantum dot.

Single electron emission is modeled by considering that each source $\alpha=1, 2$ generates in the outer edge channel the prepared state $|\varphi_\alpha,F\rangle = \int \mathrm{d}\omega \, \widetilde{\varphi}_{e,\alpha}(\omega)c^\dagger(\omega)|F\rangle$ where $c^\dagger(\omega)$ creates an electron at a given energy $\omega$ on top of the Fermi sea. Here $\widetilde{\varphi}_{e,\alpha}(\omega) $ is the single electron wavefunction generated by source $\alpha$ in the energy domain given by Eq.(\ref{eqphifour}) (we assume that the hole emission process at energy $\omega_h=\omega_e$ gives the same result as the electron case). An equivalent description of the initial state can be written as a quantum superposition in real space \cite{Ferraro2014, Wahl2014, Slobodeniuk2016} by extending the non interacting region for $x \leq 0$. In this description, the initial state is represented by the exponentially decaying wavepacket of Eq.(\ref{eqphi}) evaluated at $t=-x/v_F$. The initial wavepacket being fully located in the non-interacting region $x\leq 0$, the velocity $v_F$ is the non-interacting Fermi velocity.

\begin{figure}[hhhhhh]
\centerline{\includegraphics[width=\columnwidth]{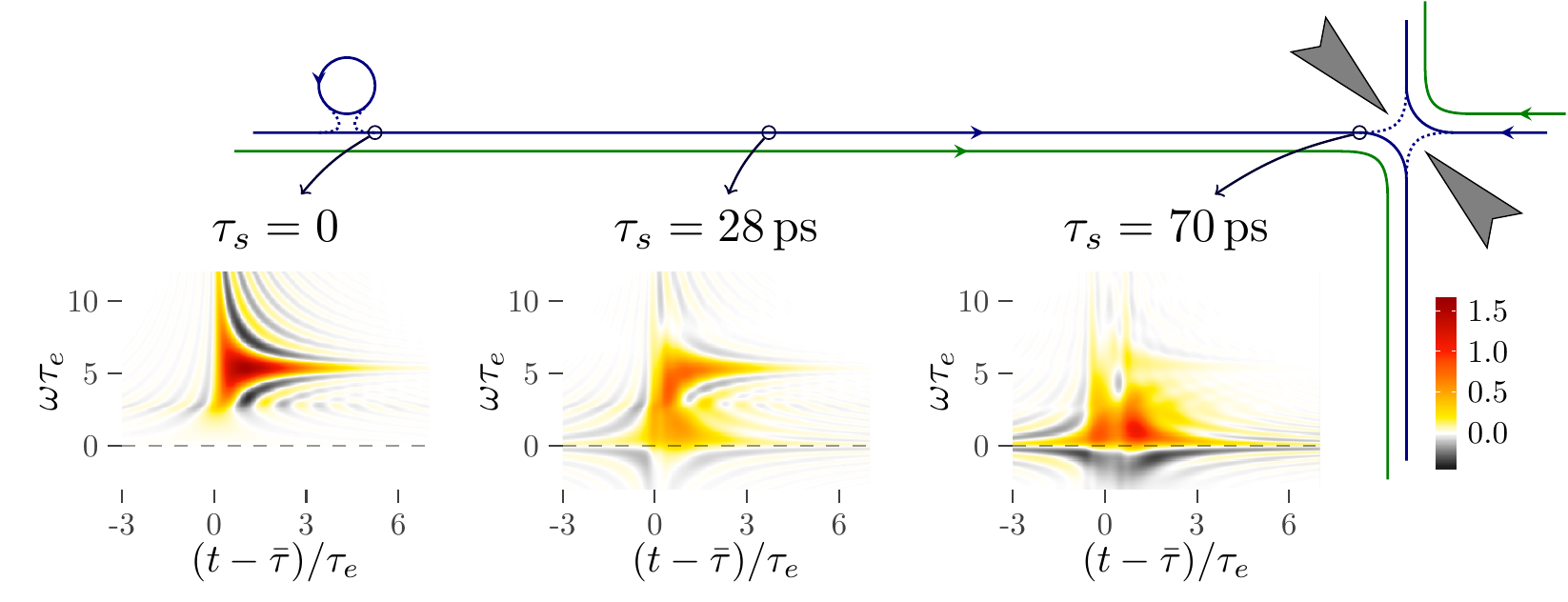}}
 \caption{ \textbf{Destruction of the elementary quasiparticle.} Wigner representations $\Delta W^{(e)}(\bar{t},\omega)$ of the excess single electron coherence at $T=0$ K for different propagation lengths $\tau_s=0$, $28$ and $70$ ps.  The time axis are shifted by time $\bar{\tau}= l/v_{\rho}$ to account for the propagation time on length $l$. For $\tau_s=0$, $\Delta W^{(e)}(\bar{t},\omega)$ represents the state emitted in the outer edge channel (blue line) described by eq. (\ref{eqphi}), with $\omega_e =0.7$ K and $\tau_e=60$ ps. For $\tau_s=28$ and $70$ ps, short range Coulomb interactions between the outer and inner (green line) edge channels are taken into account. } \label{Fig5}
\end{figure}

Starting from this initial state, analytical calculations \cite{Wahl2014, Ferraro2014} have been developed to compute the electronic coherence $\Delta \mathcal{G}^{(e)}(t,t')$ and the HOM correlations $\Delta q(\tau)$ at the output of the interaction region. Using the bosonization technique and numerical evaluation of the resulting non-perturbative expressions, we first numerically compute the excess electronic coherence function at $T=0$K in the Wigner representation \cite{Ferraro2013} $\Delta W^{(e)}(\bar{t},\omega)$, obtained from the Fourier transform of $\Delta \mathcal{G}^{(e)}(\bar{t}+ \tilde{\tau}/2,\bar{t}-\tilde{\tau}/2)$ on the time difference $\tilde{\tau}$.  $\Delta W^{(e)}(\bar{t},\omega)$ is plotted on Fig.\ref{Fig5} for increasing values of the propagation length or equivalently of $\tau_s$. The choice of the Wigner representation proves particularly useful here, first because it combines temporal and energetic aspects of single electron decoherence, but also because it allows for a classical limit, where $W^{(e)}(\bar{t},\omega)$ represents the occupation probability of states at energy $\omega$ as a function of time $\bar{t}$. This classical interpretation fails for $ W^{(e)}(\bar{t},\omega) <0 $ or $W^{(e)}(\bar{t},\omega) > 1$ \cite{Ferraro2013}.

Here, $\Delta W^{(e)}(\bar{t},\omega)$ provides a direct visualization of the evolution of the single electron wavepacket under the influence of Coulomb interaction leading to the destruction of the single electron\cite{Ferraro2014} as shown on Fig.\ref{Fig5}. For $\tau_s=0$, $\Delta W^{(e)}(\bar{t},\omega)$ corresponds to the Wigner representation of the pure single electron state (\ref{eqphi}). Along the energy axis, $\Delta W^{(e)}$ is very broad on short times (as a consequence of the Heisenberg uncertainty principle) and becomes peaked around the emission energy $\omega_e$ on a typical timescale given by $\tau_e$. Ripples of negative or above unity values of $\Delta W^{(e)}(t,\omega)$ show the non-classical nature of the single electron state. After a short propagation length, $\tau_s = 28$ ps, before the fractionalization in two pulses has occurred, energy relaxes and the spectral weight at $\omega_e$ is transferred close to the Fermi energy ($\omega_e = 0$). The non-classical ripples are also almost completely washed out. On longer propagation length $\tau_s=70$ ps, the fractionalization in two distinct pulses occurs and can be seen along the temporal axis. As two pulses of charge $e/2$ cannot correspond to a single quasiparticle excitation of the Fermi sea, collective neutral excitations are created. This can be seen on $\Delta W^{(e)}(\bar{t},\omega)$ by its negative values below the Fermi energy (corresponding to the creation of holes) compensated by an increase of $\Delta W^{(e)}(\bar{t},\omega)$ above the Fermi energy (corresponding to the creation of the same number of electrons).

As can be seen on Eq.(\ref{DeltaSHOM}), the HOM dip encodes information on the overlap of single electron coherences, which can be rewritten in term of overlap between the Wigner distributions of sources 1 and 2.:
 \begin{equation}
\Delta S_{HOM} = 4  R(1-R) e^2\int \frac{d\omega}{2 \pi} \overline{\Delta W^{(e)}_1(\bar{t},\omega) \Delta W^{(e)}_2(\bar{t},\omega)}^{\bar{t}}
 \end{equation}
 The single electron decoherence scenario represented by the evolution of the Wigner distribution $\Delta W^{(e)}(\bar{t},\omega)$ as a function of propagation length on Fig.\ref{Fig5} can thus be tested by means of HOM data versus theory comparison\cite{Wahl2014, Ferraro2014}.

\begin{figure}[hhhhhh]
\centerline{\includegraphics[width=20cm]{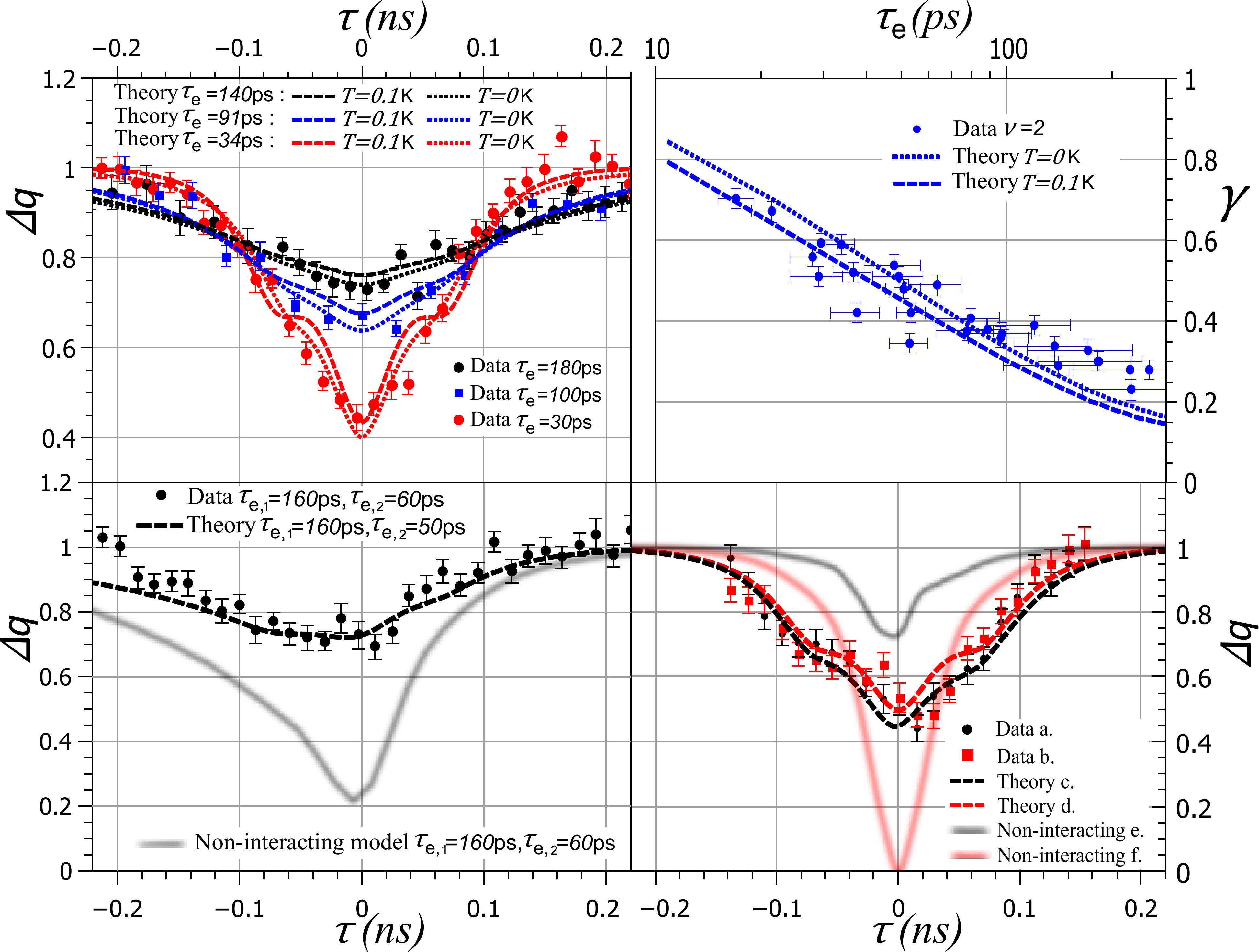}}
\caption{ \textbf{Data/model comparison.} Upper left panel, $\Delta q(\tau)$ for various emission times. Theory accounting for  Coulomb interaction is represented in dotted line ($T=0$ K) and dashed line ($T=0.1$ K). Lower left panel, $\Delta q(\tau)$ for asymmetric emission times. Theory predictions accounting for Coulomb interaction ($T=0.1$ K) are represented in dashed lines. Predictions of the non-interacting model in blurred black. Upper right panel, contrast $\gamma$ versus emission time $\tau_e$ (in log-linear scale). The dotted ($T=0$) and dashed ($T=100$ mK) lines represent theory predictions accounting for Coulomb interaction. Lower right panel, a. Data, $\tau_e=40$ ps, $400$ mK gate noise on dot 1. b. Data, $\tau_e=40$ ps without gate noise. c. Theory, $T=0.1$ K, $\omega_1 = 0.7$ K, $\omega_2=0.3$ K, $\tau_e=40$ ps. d. Theory, $T=0.1$ K, $\omega_1 = \omega_2=0.7$ K e. Non-interacting model, $\omega_1 =0.7$ K, $\omega_2=0.3$ K, $\tau_e=40$ ps. f. Non-interacting model, $\omega_1 =\omega_2=0.7$ K, $\tau_e=40$ ps.}
\label{Fig6}
\end{figure}

The upper left panel of Fig.\ref{Fig6} presents the data of the HOM traces $\Delta q(\tau)$ for various emission times $\tau_e$ together with theoretical predictions  at $T=0$ and $T=100$ mK providing an evaluation of the effect of finite temperature on single electron decoherence.  The interaction parameter is set to $\tau_s=70$ ps which is extracted from high frequency admittance measurements \cite{Bocquillon2013b} performed on a similar sample coming from the same batch and which confirmed the validity of the short range interaction model up to frequencies $f \leq 6$~GHz. The parameter  $\tau_s=70$ ps has also been successfully used to describe the charge fractionalization in Ref. \cite{Freulon2015} using the same sample as in the present work (at the same value of the magnetic field). The red, blue, and black curves represent these theoretical predictions taking $\tau_e=34$, $91$ and $147$ ps. These values agree within experimental resolution with the values of $\tau_e$ extracted from the measurements of the average current. In particular for the short time $\tau_e=34$ ps, theoretical predictions capture the broadening of the electronic wavepacket by the fractionalization process which leads to an overestimate by a factor 2 of the emission time extracted from the exponential fit of the dip (although the experimental resolution is not good enough to observe the predicted side peaks for $\tau_e=34$ ps, $T=0.1$ K). The agreement between the data and the predictions is good: once the width of the dip has been chosen to match the data, the values of the contrast also agree. Note that contrary to Fig.\ref{Fig2} where a phenomenogical description of decoherence involving two adjustable parameters was used, we are able here to fit the full HOM trace using only experimentally measured parameters (emission time $\tau_e$ and interaction strength $\tau_s$). The differences between the calculated HOM curves at different temperatures are small showing a small influence of temperature on single electron decoherence. This is explained by the electron emission energy $\hbar \omega_e >k_bT_e$.

The lower left panel of Fig.\ref{Fig6} presents the data-model comparison when the emission times of the two sources are detuned. The agreement is also very good providing the right value for the contrast of two-particle interferences contrary to the non-interacting predictions (black blurred line).

Last but not least, the lower right panel of Fig.\ref{Fig6} exhibits the most striking distinctive prediction of the interaction model: the contrast and shape of the HOM trace is almost unchanged when the emission energy of one of the two sources is varied (from $0.7$ K to $0.3$ K). This behavior is completely different from the non-interacting model predictions (black and red blurred lines) for which the contrast varies strongly from $1$ to $0.25$ when the energies are detuned by $400$ mK at $\tau_e=40$ ps. Surprisingly, in the detuned case, interactions lead to enhancement of the contrast compared to the non-interacting prediction. This restauration of indistinguishability by decoherence is a consequence of electronic
relaxation. At a quantitative level, it can be shown that at long times, the resulting single
electron coherence only depends on the shape of the initial current pulse (here encoded in
the duration $\tau_e$) and of the propagation distance but not anymore of its initial injection
energy. This erasure effect is a consequence of the entanglement of the electronic degrees
of freedom of the outer edge channel where the single electron excitation is injected with
the inner one \cite{Ferraro2014}. Quantitatively confirming this effect is a strong signature of the single
electron decoherence scenario described within the bosonization framework.

Finally, the data model comparison of the $\tau_e$ dependence of the contrast $\gamma$  can be seen on the upper right panel of Fig.\ref{Fig6} giving a coherence time $\tau_c=86$ ps ($T=100$ mK) and $\tau_c=104$ ps ($T=0$ K) close to the data ($\tau_c=98$ ps at $\nu=2$). The overall agreement is good even if, for long escape times, the data tends to accumulate above the theoretical predictions. However, for $\tau_e \geq 180$ ps, $\tau_e$  cannot be neglected compared to the drive half-period and the probability $P$ of single charge emission starts to decrease ($P\leq 0.9$) \cite{Mahe2010, Albert2010}. The comparison with perfect single electron states thus ceases to be valid in this long emission time limit.

\section{Conclusion}

To conclude, we have analyzed the coherence of single electron states propagating along a one-dimensional edge channel using HOM interferometry. We observe a strong reduction of the HOM contrast when the width of the emitted single electron wavepackets is increased from which a coherence time $\tau_c \approx 100$ ps (at $\nu=2$) can be extracted. Our results are in quantitative agreement with the Coulomb interaction induced decoherence along propagation providing the first direct evidence of the destruction scenario of a single quasiparticle in a one dimensional conductor. The outcome of this study extends, beyond charge propagation in conductors, to a large variety of low-dimensional systems where the Luttinger paradigm is relevant \cite{Kollath2005, Angelakis2011, Jurcevic2015}. For example, those decoherence scenarii could be studied as a function of interaction strength in low dimensional cold atomic systems where single elementary excitations can now be manipulated \cite{Fukuhara2013} and experimental resolutions now reach the single atom scale \cite{Haller2015}.

\section*{Acknowledgments}
 This work is supported by the ERC consolidator grant 'EQuO' and the ANR grant '1shot reloaded', ANR-14-CE32-0017. The development of the HEMTs used for cryogenic readout electronics in this experiment was supported in part by the European FP7 space project CESAR grant No. 263455.

\end{document}